\documentstyle[12pt]{article}
\renewcommand{\thefootnote}{\fnsymbol{footnote}}

\begin{document}\begin{titlepage} \vspace{1.3cm}

\rightline{DFPD96/TH/50}

\rightline{\tt hep-th/9610204} 

\rightline{Phys. Rev. Lett. {\bf 78} (1997) 1412.}

\vspace{1.5cm}

\centerline{\large{\bf MODULAR INVARIANCE AND STRUCTURE OF THE}} \vspace{0.6cm}

\centerline{\large{\bf EXACT WILSONIAN ACTION OF N=2 SYM}\footnote[3]{Work 
supported by the European Commission TMR programme ERBFMRX--CT96--0045.}}

\vspace{1.5cm} {\centerline{\sc MARCO MATONE}} \vspace{0.8cm}

\centerline{\it Department of Physics ``G. Galilei'' - Istituto Nazionale di 
Fisica Nucleare} \centerline{\it University of Padova} 
\centerline{\it Via Marzolo, 8 - 35131 Padova, Italy}

\centerline{matone@padova.infn.it} \vspace{2cm}

\centerline{\bf ABSTRACT} \vspace{0.6cm}

\noindent
We construct modular invariants on ${\cal M}_{SU(2)}$, the moduli space of 
quantum vacua of $N=2$ SYM with gauge group $SU(2)$. We also introduce the 
nonchiral function ${\cal K}(A,\bar A)=2\alpha \pi e^{\varphi_{SW}-
\varphi/2}$, where $e^{\varphi_{SW}}$ is the Seiberg--Witten metric, 
$e^\varphi$ is the Poincar\'e metric on ${\cal M}_{SU(2)}$ and $\alpha$ is a 
regularization scheme--dependent constant. It turns out that ${\cal K}(A,\bar 
A)$ has all the expected properties of the next to leading term in the 
Wilsonian effective action ${\cal S}[A,\bar A]$ whose modular properties are 
considered in the framework of the dimensional regularization.

\end{titlepage}\newpage\setcounter{footnote}{0} 
\renewcommand{\thefootnote}{\arabic{footnote}}

\noindent
The exact results about $N=2$ SUSY Yang--Mills obtained by Seiberg and Witten 
\cite{SW1} concern the low--energy Wilsonian effective action with at most two 
derivatives and four fermions. In the $SU(2)$ case, the $u$--moduli space of 
quantum vacua is ${\cal M}_{SU(2)}$, the Riemann sphere with punctures at
$u=\infty$, $u=\pm\Lambda^2$. In \cite{BMT}  
results in \cite{SW1} have been derived from first principles
\cite{mm,DKM1FTDKM2HW}. In particular, the $T^2$ symmetry $u(\tau+2)=u(\tau)$, 
which rigorously follows from the asymptotic analysis together with the 
relation \cite{mm} 
\begin{equation}
u=\pi i ({\cal F}-a\partial_a{\cal F}/2), \label{81}\end{equation}
and the fact that
\begin{equation}
\overline{u(\tau)}=u(-\bar\tau), \qquad u(\tau+1)=-u(\tau),
\label{hfgt}\end{equation}
uniquely fix the monodromy group $\Gamma$ to be $\Gamma(2)$. The basic 
observation is that, for real values of $u$ we have the symmetry  ${u(\tau)}=
u(-\bar\tau)$ which essentially fixes $\Gamma$. The reason is that by (\ref{81})
and ${\rm Im}\, \tau>0$ (except for the singularities where ${\rm Im}\, 
\tau=0$), $u$ can be seen as uniformizing coordinate. Therefore, 
${\cal M}_{SU(2)}\cong H/\Gamma$ where $H$ is the upper half plane (the 
$\tau$--moduli space; see \cite{m2}). This is equivalent to 
$u(\gamma\cdot \tau)=u(\tau)$ with $\gamma\in \Gamma$. It follows that there 
are curves ${\cal C}$ in the fundamental domains in $H$ such that for $\tau\in 
{\cal C}$ one has $\gamma\cdot \tau=-\bar\tau$. This reasoning together with a 
proper use of $u(\tau+1)=-u(\tau)$ essentially implies the results 
in \cite{BMT}.

In \cite{SW1} it has been emphasized that the metric
\begin{equation}
ds^2={\rm Im}\,\left({\partial^2_a{\cal F}}\right)|da|^2=
|\partial_ua|^2{\rm Im}\,\left({\partial^2_a{\cal F}}\right)
|du|^2, \label{15}\end{equation}
is at heart of the physics. The natural framework to 
investigate its properties is uniformization theory \cite{mm,m2,BMT}.

In this Letter we use basic geometrical structures of ${\cal M}_{SU(2)}$ to 
derive a modular invariant quantity which fulfills all the expected properties 
of the next to leading term in the Abelian Wilsonian effective 
action \cite{Henningson,deWitGrisaruRocek,Yung,FordSachs}.

Let us now recall the metric introduced in \cite{m2}. Let $H=\{w|{\rm Im}\, 
w>0\}$ be the upper half plane endowed with the 
Poincar\'e metric $ds^2_P=({\rm Im}\, w)^{-2}|dw|^2$. Since $\tau=\partial_a^2 
{\cal F}$ is the inverse of the map uniformizing ${\cal M}_{SU(2)}$,
it follows that the positive definite  metric
\begin{equation}
ds_P^2={\left|{\partial^3_a{\cal F}}\right|^2\over\left({\rm Im}\, 
\tau\right)^2}|da|^2={\left|\partial_u \tau\right|^2\over\left({\rm Im}\,
\tau\right)^2}|du|^2=e^{\varphi} |du|^2,
\label{16bb}\end{equation}
is the Poincar\'e metric on ${\cal M}_{SU(2)}$. This implies that $\varphi$ 
satisfies the Liouville equation $\varphi_{u\bar u}=e^{\varphi}/2$. 

We now show that the Seiberg--Witten metric 
\begin{equation} e^{\varphi_{SW}}=|a'|^2 {\rm Im}\,\tau,
\label{gdft}\end{equation}
where $'\equiv\partial_u$, can be written in terms of the Poincar\'e metric 
$e^\varphi$. To this end we first summarize a few facts.
A crucial role in the theory is played by the Picard--Fuchs
equations \cite{KLT,CerDaFe}; in particular in the $SU(2)$ 
case we have the (reduced) uniformizing equation \cite{KLT,mm}
\begin{equation}
\left[4(u^2-\Lambda^4)\partial_u^2+1\right]\psi=0,
\label{2}\end{equation}
satisfied by $a_D$ and $a$, implying that 
\begin{equation}
(\Lambda^4-{\cal G}^2)\partial_a^2{\cal G}+{a\over 4}\left(\partial_a
{\cal G}\right)^3=0,
\label{1}\end{equation}
where $u={\cal G}(a)$.
By (\ref{81}) and (\ref{1}) we have \cite{m2}
\begin{equation}
\partial_a^3 {\cal F}={\pi^2\left(a \partial_a^2{\cal F}-\partial_a{\cal F}
\right)^3\over 16\left[{\Lambda^4}+\pi^2\left({\cal F}-{a}\partial_a{\cal F}/2
\right)^2\right]}.
\label{gdtf}\end{equation}
Furthermore, by \cite{SW1}
\begin{equation}
a_D=\partial_a{\cal F}={\sqrt 2\over \pi}\int_{\Lambda^2}^u {dx \sqrt{x-u}\over 
\sqrt{x^2-{\Lambda^4}}},\qquad a={\sqrt 2\over\pi}
\int_{-{\Lambda^2}}^{\Lambda^2}{dx \sqrt{x-u}\over \sqrt{x^2-{\Lambda^4}}},
\label{14}\end{equation}
we have $a(-{\Lambda^2})= -i4\Lambda/\pi$ and $a({\Lambda^2})=
4\Lambda/\pi$. It follows that the initial conditions for the second--order 
differential equation (\ref{1}) are ${\cal G}(-i4{\Lambda}/\pi)
=-{\Lambda^2}$ and ${\cal G}(4{\Lambda}/\pi)={\Lambda^2}$ and by (\ref{81})
\begin{equation}
{\cal F}(a)={2i\over \pi}a^2\int^a_{4\Lambda/\pi}db {\cal G}(b)b^{-3}
-{\pi i \over 16} a^2. \label{aaiqwnd}\end{equation}

Now observe that by (\ref{gdtf})
\begin{equation} \partial_u\tau={1\over 2\pi i {a'}^2(\Lambda^4-u^2)}.
\label{idhpo}\end{equation}
Since $e^\varphi={\left|{\partial_u \tau }\right|^2/\left({\rm Im}\, 
\tau\right)^2}$, we can rewrite Eq.(\ref{idhpo}) as
\begin{equation} e^{-\varphi/2}= 2\pi |a'|^2 |u^2-\Lambda^4|{\rm Im}\, \tau.
\label{come}\end{equation}
We stress that, as observed in \cite{mm} for the term $|u^2-\Lambda^4|$
in the uniformizing equation, also  in (\ref{come})  it should be 
considered as a $(-3/2,-3/2)$ differential on ${\cal M}_{SU(2)}$. This ensures
the covariance of (\ref{come}). By (\ref{gdft}) we have 
\begin{equation} e^{\varphi_{SW}}= {e^{-\varphi/2}\over 2\pi|u^2-\Lambda^4|}.
\label{comepuo}\end{equation}
Now observe that by (\ref{81}) we have 
\begin{equation} a'={2\over \pi i a(\hat \tau -\tau)},
\label{wehave3}\end{equation}
where $\hat\tau=a_D/a$. Therefore, (\ref{come}) is equivalent to
\begin{equation}
e^{-\varphi/2}={8\over \pi} {|u^2-\Lambda^4|\over|a|^{2} |\hat \tau -\tau|^{2}}
{\rm Im}\,\tau. \label{gdtfy}\end{equation}

In \cite{Henningson} it has been shown that the next to leading term which 
contributes to the Abelian Wilsonian effective action ${\cal S}[A,\bar A]$ is 
$\int d^4x d^4\theta d^4\bar \theta {\cal K}(A,\bar A)$ where ${\cal K}(A,
\bar A)$ is a modular invariant real analytic function of the $N=2$ $U(1)$ 
vector multiplet $A$. Therefore, up to this order ${\cal S}[A,\bar A]$
has the form $\int d^4x d^4\theta d^4\bar \theta {\cal K}(A,\bar A)
+{1\over 4\pi}{\rm Im}\,\left(\int d^4x d^4\theta {\cal F}(A)\right)$. 
In Ref.\cite{deWitGrisaruRocek} De Wit, Grisaru and Ro\v{c}ek were able to 
prove that asymptotically  
\begin{equation}
{\cal K}(A,\bar A)\sim c\ln {A\over \Lambda}\ln {\bar A\over\Lambda},
\label{hasthestr}\end{equation}
where $c$ is a constant. Furthermore, the one--instanton contribution to 
${\cal K}$ has been obtained by Yung \cite{Yung}. We observe that 
one should expect that the logarithmic singularities at $u\to \infty$, such as 
the asymptotic behavior in (\ref{hasthestr}), be the unique singularities in 
${\cal K}$ as possible singularities at $u\ne\infty$
will spoil the physical meaning of the quantum moduli 
space. In particular, ${\cal K}$ should be regular where monopoles 
or dyons become massless (see also \cite{Yung}). Actually, it seems that the 
only possible way in order to not change the physical picture in \cite{SW1}
is that ${\cal K}$ be vanishing at these points. Let us illustrate 
this aspect by recalling that as a crucial property of the $N=2$ SYM 
path--integral measure, there is a dual version of the theory where the fields 
are the $S$ transformed of the original ones. In particular, the dual 
effective coupling constant is
\begin{equation} \tau_D=-{1\over \tau}.
\label{laemieabbastanza}\end{equation}
Now observe that the $\Gamma(2)$ symmetry can be also interpreted in the 
following way. Let us schematically denote by $I$ and $II$ the original theory 
and its dual respectively
\begin{equation} S\cdot I=II.
\label{cdotte}\end{equation}
As $T^2$ is a symmetry of $I$ theory, we have $T^2\cdot I=I$. On the other 
hand, the properties of the $N=2$ measure imply that the asymptotic analysis 
still holds for the dual theory $II$, that is $T^2\cdot II=II$. It follows that
\begin{equation} S^{-1}T^2S\cdot I=S^{-1}T^2\cdot II=S^{-1}\cdot II=I,
\label{andaeriandatrick}\end{equation}
implying that besides $T^2$ also $S^{-1}T^2S$ is in the symmetry group.
Repeating the steps in (\ref{andaeriandatrick}) we generate all
$\Gamma(2)$, which is the symmetry group of $I$ and of its dual version $II$. 
This fact suggests that the higher--order terms in ${\cal S}[A,\bar A]$ 
should have a structure such that the asymptotic behavior (\ref{hasthestr}) 
still holds for the dual theory $II$. In particular, the  asymptotic behavior
$\tau\sim {2i\over \pi}\ln a$ is reproduced in the dual theory as $\tau_D\sim 
-{i\over \pi}\ln a_D$. Since these points are in the spectrum,
it follows by (\ref{laemieabbastanza}) that $\tau=0$ for some $u$. This is the 
$u=\Lambda^2$ point. A similar property should still hold for the
higher--order terms in ${\cal S}[A,\bar A]$. Therefore, in order to 
preserve the relation $S\cdot I=II$, it seems that ${\cal K}(a,\bar a)$
should vanish at the puncture $u=\Lambda^2$. Furthermore, modular invariance 
of ${\cal K}$ together with the relation $u(\tau+1)=-u(\tau)$, implies
that ${\cal K}$ is invariant under  $u\to -u$. It follows that ${\cal K}$ 
should vanish at $u=\pm \Lambda^2$. This argument may be better formulated by 
requiring that the $T$ transformation $\tau\to\tau+1$, which corresponds to 
$a\to e^{-\pi i/2}a$ (i.e. $u\to -u$), be an invariance of ${\cal K}$, namely
\begin{equation} {\cal K}(e^{-\pi i/2}A,e^{\pi i/2}\bar A)={\cal K}(A,\bar A).
\label{invs}\end{equation}

We now show that there is a natural choice for ${\cal K}(A,\bar A)$
which fulfills all the above conditions. Namely, we propose that
\begin{equation}
{\cal K}(A,\bar A)=\alpha{e^{-\varphi({\cal G}(A),\overline{{\cal G}(A)})}
\over |{\cal G}^2(A) - \Lambda^4|}, \label{ganzo}\end{equation}
where $\alpha$ will be fixed using the one--instanton calculation 
in \cite{Yung}. By (\ref{comepuo}) it follows that ${\cal K}(A,\bar A)$
can be also written in the form ${\cal K}(A,\bar A)=4\alpha\pi^2
{e^{2\varphi_{SW}}|{\cal G}^2(A) - \Lambda^4|}$ or
\begin{equation}
{\cal K}(A,\bar A)=2\alpha\pi e^{\varphi_{SW}({\cal G}(A),\overline{{\cal 
G}(A)})-{\varphi\over 2}({\cal G}(A),\overline{{\cal G}(A)})}.
\label{ganzotris}\end{equation}

In the following we will show that the solution (\ref{ganzo})
has the following properties

\vspace{0.5cm}

\noindent {\bf 1.} It is modular invariant.

\vspace{0.5cm}

\noindent {\bf 2.} Asymptotically ${\cal K}(A,\bar A)\sim 2\alpha\ln 
{A/\Lambda}\ln {\bar A/\Lambda}$.

\vspace{0.5cm}

\noindent {\bf 3.} The above is the unique singularity of ${\cal K}$.

\vspace{0.5cm}

\noindent {\bf 4.} The zeroes of ${\cal K}$ 
are precisely at the punctures.

\vspace{0.5cm}

\noindent
{\bf 5.} Besides the logarithmic terms the asymptotic expansion of ${\cal K}
(A,\bar A)$ contains terms like $(A/\Lambda)^{-4j}(\bar A/\Lambda)^{-4k}$ as 
expected from the instanton contributions (see also \cite{Yung}).

\vspace{0.5cm}

We now make a few remarks concerning point $\bf 1.$ Strictly speaking, a 
function $G(A,\bar A)$ is said to be modular invariant if
$G(\gamma \cdot A,\gamma \cdot \bar A)=G(A,\bar A)$, $\gamma\in SL(2,{\bf Z})$.
The function ${\cal K}(A,\bar A)$ has the invariance $T\circ {\cal K}(A,\bar A)
={\cal K}(A,\bar A)$ and $S\circ {\cal K}(A,\bar A)={\cal K}(A,\bar A)$. 
Whereas in the first case there is not any change in the functional structure 
of ${\cal K}$, in the case of the $S$--action one has  $S\circ {\cal K}(A,\bar 
A)={\cal K}_D(A_D,\bar A_D)$ \cite{Henningson}. A similar situation arises in 
the case of $u={\cal G}(a)$. Since under a $SL(2,{\bf C})$ transformation 
${\cal F}(a)$ and $aa_D/2$ have the same transformation properties, it follows 
that under $SL(2,{\bf C})$ one has $\tilde {\cal G}(\tilde a)={\cal G}(a)$. 
However, observe that the $T$--action $\tau\to \tilde\tau=\tau+1$ is generated 
by a phase change of $a$. In particular, under $a\to\tilde a=e^{\pi i n/2}a$, 
we have ${\cal F}(\tilde a)=e^{\pi i n}{\cal F}(a)-e^{\pi i n}na^2/2$, $\tilde 
a_D=\partial_{\tilde a}{\cal F}(\tilde a)=e^{\pi i n/2}a_D-n e^{\pi i n/2} a$, 
$\tau\to \tilde\tau=\tau-n$. Concerning the $S$--action we have ${\cal G}_D
(a_D)=S \circ {\cal G}(a)={\cal G}(a)$. Therefore, since ${\cal K}(A,\bar A)$ 
in (\ref{ganzo}) is expressed in terms of ${\cal G}(A)$ and  $\overline{{\cal 
G}(A)}$, it follows that ${\cal K}_D(A_D,\bar A_D)={\cal K}(A,\bar A)$ as it 
should be \cite{Henningson}. Under the transformation $a\to \tilde a=e^{-\pi i 
/2}a$, corresponding to the $T$--action $\tau\to\tilde \tau=\tau +1$, we have 
${\cal G}(\tilde a)=-{\cal G}(a)=-\tilde {\cal G}(\tilde a)$. On the other hand,
by the ${\bf Z}_2$ automorphism of ${\cal M}_{SU(2)}$, we 
have $e^{\varphi(-u,-\bar u)}=e^{\varphi(u,\bar u)}$, so that by (\ref{ganzo})
${\cal K}$ satisfies Eq.(\ref{invs}). Finally, we observe that 
by the $\Gamma(2)$ symmetry ${\cal G}(\gamma\cdot A)={\cal G}(A)$,
we have ${\cal K}(\gamma\cdot A,\gamma\cdot \bar A)={\cal K}(A,\bar A)$.

In order to consider point $\bf 2.$ we first write down the asymptotic 
expansions
\begin{equation}
{\cal F}=a^2\left[{i\over \pi}\ln {a\over \Lambda}+\sum_{k=0}^\infty{\cal F}_k
\left({a\over \Lambda}\right)^{-4k}\right], \label{hfgty1}\end{equation}
\begin{equation}
\tau={2i\over\pi}\ln{a\over\Lambda}+{3i\over\pi}+\sum_{k=0}^\infty {\cal F}_k
(1-4k)(2-4k)\left({a\over \Lambda}\right)^{-4k}. \label{hfgty3}\end{equation}
By (\ref{81}) and (\ref{hfgty1}) it follows that the asymptotic expansion for 
$u={\cal G}(a)$ is 
\begin{equation}
{\cal G}(a)=a^2\sum_{k=0}^\infty {\cal G}_k \left({a\over\Lambda}\right)^{-4k},
\qquad {\cal G}_0={1\over 2}, \label{abas}\end{equation}
where  ${\cal G}_k=2 \pi i k{\cal F}_k$, for $k>0$. Concerning the instanton 
contributions ${\cal F}_k$, $k>0$, these are determined by the recursion 
relations \cite{mm}
$$ {\cal G}_{n+1}={1\over 8{\cal G}_0^2(n+1)^2}\cdot $$
\begin{equation}
\cdot\left[(2n-1)(4n-1){\cal G}_n+2\sum_{j=0}^{n-1}{\cal G}_{n-j}
\left({\cal G}_{j+1}{\cal G}_0c(j,n)-
\sum_{k=0}^{j+1}{\cal G}_{j+1-k}{\cal G}_{k}
d(j,k,n)\right)\right],\; n\ge 0, \label{recursion2}\end{equation}
where $c(j,n)=2j(n-j-1)+n-1$, $d(j,k,n)=[2(n-j)-1][2n-3j-1+2k(j-k+1)]$.
To evaluate ${\cal F}_0$ we observe that for $u\to\infty$, Eq.(\ref{14}) yields
$a_D\sim {i\over\pi}\sqrt{2u}(\ln u/\Lambda^2 +\ln 8/e^2)$, whose $a$ term 
is, by (\ref{abas}), ${i\over\pi}a\ln4/e^2\Lambda^2$. Comparing with the 
$a$ term ${i\over \pi}a(\ln e/\Lambda^2-2\pi i {\cal F}_0)$ in the 
asymptotic expansion of $a_D$ which follows by (\ref{hfgty1}), we obtain
${\cal F}_0={i\over 2\pi}\ln 4/e^3$.

By (\ref{gdtfy}) and (\ref{ganzo}), we can rewrite ${\cal K}(A,\bar A)$ in 
the form
\begin{equation}
{\cal K}(A,\bar A)={64\alpha\over \pi^2}{|{\cal G}^2(A)-\Lambda^4|({\rm Im}\,
\tau(A))^2\over |A|^{4}|\hat \tau(A) -\tau(A)|^4}. \label{dojp}\end{equation}
Therefore, by (\ref{hfgty1})(\ref{hfgty3})(\ref{abas}) and (\ref{dojp}) we 
have the asymptotic expansion
$${\cal K}(A,\bar A)={64 \alpha\over \pi^2}\left\{{1\over \pi}\ln {A\over 
\Lambda}{\bar A\over \Lambda}+{3\over \pi}+{1\over 2i}\sum_{k=0}^\infty
{\cal F}_k (1-4k)(2-4k)\left[\left({A\over \Lambda}\right)^{-4k}+ 
\left({\bar A\over \Lambda}\right)^{-4k}\right]\right\}^2\cdot $$
\begin{equation}
\cdot{\left|\left[\sum_{k=0}^\infty {\cal G}_k\left({A\over \Lambda}
\right)^{-4k}\right]^2-\left({A\over \Lambda}\right)^{-4}\right|\over
\left| {2i\over \pi}+\sum_{k=0}^\infty
{\cal F}_k 4k (4k-2)\left({A\over \Lambda}\right)^{-4k}\right|^4}.
\label{hfgyt}\end{equation} Extracting the leading term we get
\begin{equation}
{\cal K}(A,\bar A)\sim 2\alpha\ln {A\over \Lambda}\ln {\bar A\over \Lambda},
\label{accordingto}\end{equation} which reproduces the expected behavior.

Let us now consider point $\bf 3.$ above. To show that ${e^{-\varphi}\over|u^2-
\Lambda^4|}$ has not other divergences outside $u=\infty$, it is useful to 
consider the form 
\begin{equation}
{\cal K}(a,\bar a)= 4\alpha\pi^2 |a'|^4 |u^2-\Lambda^4|({\rm Im}\, \tau)^2,
\label{comebis}\end{equation}
and to notice that by (\ref{14}), $a'(u)$ is logarithmically divergent for $u
\to \pm \Lambda^2$. Therefore, by (\ref{comebis}) ${\cal K}$ vanishes at $u=\pm
\Lambda^2$. It follows that ${\cal K}$ is everywhere finite except that for the
asymptotic divergence (\ref{hfgyt}). Furthermore, since the only zeros 
of the Poincar\'e metric 
$e^\varphi$ are at the puncture at $u=\infty$, where 
$\varphi\sim -2\ln(|u|\ln|u|)$, whereas the unique divergences come from the 
punctures at $u=\pm\Lambda^2$, where $\varphi\sim -2\ln(|u\mp\Lambda^2|\ln|u
\mp\Lambda^2|)$, it follows that ${\cal K}(a,\bar a)$ has zeros at $a=-i4
\Lambda/\pi$, $a=4\Lambda/\pi$ (and their $\Gamma(2)$ transformed). Finally, 
note that point $\bf 5.$ follows explicitly from the asymptotic expansion 
(\ref{hfgyt}). 

Let us comment on the possible higher--order terms in ${\cal S}$. 
According to \cite{Henningson}, besides ${\cal K}$, the possible 
higher--order terms do not seem to be modular invariants. On the other hand, 
as $u(\gamma\cdot\tau)=u(\tau)$, $\gamma\in \Gamma(2)$, possible non invariant 
terms should imply that at the same point of the moduli space of quantum vacua 
${\cal M}_{SU(2)}$ there are inequivalent theories. This would break the highly
symmetric structure coming from the lower--order part of the action. In this 
context one should investigate whether the nice observation in \cite{FordSachs}
that the $S$ duality transformation extends to the full effective action, 
implies that actually the higher--order terms in ${\cal S}$ are 
modular invariants. In order to better illustrate this point, we consider the 
following decomposition ${\cal S}[A,\bar A]=\widehat {\cal S}[A,\bar A]+
{1\over 4\pi}{\rm Im}\,\left(\int d^4x d^4\theta {\cal F}(A)\right)$. From the 
definition given in \cite{Henningson}, it follows that $\widehat {\cal S}[A,
\bar A]$ is $T$ invariant. This happens also for the $T$--action on the 
higher--order part of the dual theory $\widehat {\cal S}_D[A_D,\bar A_D]$. So 
that, the fact that $S^2={\bf I}$ and $(ST)^3={\bf I}$, indicates that
$\widehat{\cal S}_D[A_D,\bar A_D]= \widehat{\cal S}[A,\bar A]$. On the other 
hand, it was argued in \cite{Henningson} that only
${\cal K}$ is modular invariant. 
We consider two possibilities to further investigate this
aspect. The first one is that there are not higher--order terms in ${\cal S}$ 
besides ${\cal K}$. This would imply that
\begin{equation}
{\cal S}[A,\bar A]=2\alpha\pi\int d^4xd^4\theta d^4\bar\theta e^{\varphi_{SW}({
\cal G}(A),\overline{{\cal G}(A)})-{\varphi\over 2}({\cal G}(A),\overline{{\cal
G}(A)})}+{1\over 4\pi}{\rm Im}\,\left(\int d^4x d^4\theta {\cal F}(A)\right).
\label{ijdh}\end{equation}
Another possibility is that the terms coming from the action of $\delta/
\delta A_D$ on the dual prepotential, and which break modular invariance (see 
\cite{Henningson} for details), are actually vanishing. In particular, whereas 
when all functional derivatives act on the exponentiated dual prepotential 
there is not any breaking of modular invariance, the other contributions have 
the effect of modifying the structure of $\widehat{\cal S}[A,\bar A]$. In other
words, forgetting the functional derivative in the right-hand side (rhs) of
$$
i^n{\delta^n\over\delta A_D^n}\exp\left[{-i\int d^4x d^4\theta{\cal F}_D(A_D)}
\right]=\exp\left[{-i \int d^4x d^4\theta  {\cal F}_D(A_D)}\right]
\left(A+i{\delta \over \delta  A_D}\right)^{n-1}A,$$
would imply that $\widehat{\cal S}$ is modular invariant. On the other hand, 
because of the functional derivative in the rhs, the above expression contains 
``$\delta(0)$'' terms. A possible way to take care of these infinities is to 
use dimensional regularization where, as well known, the 
``$\delta(0)$'' terms vanish for dimensional reasons (see for example 
\cite{Collins}).

Let us now fix the constant $\alpha$. To this end we use the one--instanton 
calculation in \cite{Yung}
\begin{equation}
{\cal K}_I(A,\bar A)= {1\over 32\pi^2}\left({A\over \Lambda}\right)^{-4}
\ln {A\over \Lambda} {\bar A\over \Lambda}+c.c.,\label{oneYung}\end{equation}
whose coefficient is $1/4$ that in
\cite{Yung}. This is a consequence of the fact that our $\Lambda$ is $\sqrt 2$ 
times the scale considered in \cite{Yung} where the Pauli--Villars 
regularization scheme was chosen.
By (\ref{hfgyt}) the coefficient of the $(A/\Lambda)^{-4}\ln {A\over
\Lambda}{\bar A\over \Lambda}$ term in the asymptotic expansion of 
${\cal K}(A,\bar A)$ is
 $2\alpha\left\{-3\pi i {\cal F}_1+\left(3-2\pi i {\cal F}_0\right)
\left[2({\cal G}_1-1)+{8\pi i}{\cal F}_1\right]\right\}=-2\alpha\left({3\over 8}
+\ln 2\right)$,
where we used the fact that ${\cal G}_0=1/2$, ${\cal F}_0={i\over 2\pi}\ln 
4/e^3$ and ${\cal G}_1=2\pi i {\cal F}_1=1/4$. Comparing 
$-2\alpha\left({3\over 8}+
\ln 2\right)$ with the coefficient in (\ref{oneYung}) we obtain
\begin{equation}\alpha=-{1\over 8(3+8\ln 2)\pi^2}.\label{alphaa}\end{equation}

An interesting point concerning the structure of the logarithmic terms in 
(\ref{hfgyt}) is that besides $\ln (A/\Lambda) \ln (\bar A/\Lambda)$ and 
$[(A/\Lambda)^{-4}+(\bar A/\Lambda)^{-4})]\ln (A \bar A/\Lambda^2)$, 
considered in \cite{deWitGrisaruRocek,Yung}, there are the terms
$\ln^2 A/\Lambda$ and $\ln^2\bar A/\Lambda$.
The fact that these should appear in the asymptotic expansion of ${\cal 
K}$ can be also seen by a simple modular invariance argument. Namely, 
whereas $(A/\Lambda)^{-4}\ln (A \bar A/\Lambda^2)$ is invariant under 
$A\to e^{-\pi i/2}A$, we have
$2\ln (e^{-\pi i/2}A/\Lambda) \ln (e^{\pi i /2}\bar A/\Lambda)=2\ln (A/
\Lambda) \ln (\bar A/\Lambda)+{\pi i}(\ln A/\Lambda-\ln \bar A/\Lambda)
+{\pi^2/ 2}$,
so that in order to satisfy (\ref{invs}) one needs more logarithmic terms.
It is easy to see that these should be of the form $(\ln^2A/\Lambda+\ln^2
\bar A/\Lambda)$. We observe that (\ref{hfgyt}) also passes this test.  
We also note that once one makes the natural choice of using
$({\rm Im}\,\tau)^2$ to reproduce both $\ln{A\over \Lambda}{\bar A\over 
\Lambda}$ and $\ln^2{A\over \Lambda}{\bar A\over \Lambda}$, then modular 
invariance essentially fixes ${\cal K}$. In this context it 
is crucial that $|A|^{-4}|\hat\tau -\tau|^{-4}$ has the same 
transformation properties of $({\rm Im}\,\tau)^2$ and, as requested by 
(\ref{hasthestr}) and (\ref{oneYung}), cancels the global 
$|A|^4$ factor in the asymptotic expansion of $|{\cal G}^2(A) - \Lambda^4|$
(see (\ref{hfgyt})).

In conclusion, we observe that similar investigations can be extended to more 
general cases. 
For example, in the $SU(3)$ case the uniformization of the quantum moduli space
has been considered in \cite{BOMA}. Also, some consequences concerning 
the Wilsonian renormalization group equation should be further 
investigated \cite{BOMA1}\cite{DHOKERKRICHEVERPHONG}.

It is a pleasure to thank D. Bellisai, G. Bonelli, F. Fucito, 
P.A. Marchetti, M. Tonin and G. Travaglini for discussions.

\end{document}